\documentclass[letterpaper]{article}
\usepackage{spconf,amsmath,epsfig}
\usepackage{cite}
\usepackage{url}
\usepackage{color,soul}
\usepackage{flushend}

\let\OLDthebibliography\thebibliography
\renewcommand\thebibliography[1]{
  \OLDthebibliography{#1}
  \setlength{\parskip}{0pt}
  \setlength{\itemsep}{0pt plus 0.3ex}
}

\newcommand{\smalllskip}[0]{\vspace{1pt}}

\newcommand{\figref}[2][{}]{{\figurename~\ref{#2}#1}}

\begin{document}\sloppy

% Example definitions.
% --------------------
\def\x{{\mathbf x}}
\def\L{{\cal L}}

\ninept

% Title.
% ------
\title{A Case Study on Video Color Transfer: Exploring User Motivations, Expectations, and Satisfaction}
%
% Single address.
% ---------------
\name{Mair\'ead Grogan*, Emin Zerman*, Gareth W. Young*, Aljosa Smolic\thanks{This publication has emanated from research conducted with the financial support of Science Foundation Ireland (SFI) under the Grant Number 15/RP/27760.}\thanks{Mair\'ead Grogan was with V-SENSE, School of Computer Science, Trinity College Dublin, Dublin, Ireland at the time of writing this preprint.}\thanks{*These authors contributed equally.}}
\address{V-SENSE, School of Computer Science, Trinity College Dublin, Dublin, Ireland}

\maketitle

\begin{abstract}
Multimedia and creativity software products are being used to edit and control various elements of creative media practices. These days, the technical affordances of mobile multimedia devices and the advent of high-speed 5G internet access mean that these abilities are simpler and more readily available to be harnessed by mobile applications. In this paper, using a prototype application, we discuss how potential users of such technology are motivated to use a video recoloring application and explore the role that user expectation and satisfaction play in this process. By exploring this topic and focusing on the human-computer interaction, we found that color transfer interactions are driven by several intrinsic motivations and that user expectations and satisfaction ratings can be maintained via clear visualizations of the processes to be undertaken. Furthermore, we reveal the specific language that users use to communicate video recoloring when regarding user motivations, expectations, and satisfaction. This research provides important information for developers of state-of-art recoloring processes and contributes to dialogues surrounding the users of mobile multimedia technology in practice.
\end{abstract}
%

% ================================
%         INTRODUCTION
% ================================
\section{Introduction}
\label{sec:intro}
In recent years there has been a growing popularity in image and video sharing via mobile applications, vastly increasing the number of casual creatives that are editing mobile multimedia content. This has resulted in a shift from more advanced, professional tools to simpler, more accessible ones~\cite{Isenberg16}. As a result, media recoloring tools that were originally assessed for usability with desktop technology and professional users in mind, must also now consider the needs of more mobile creatives~\cite{zhang2005challenges, coursaris2011meta}. 

This raises many new and interesting questions for the research and development of new software, such as: \textit{What motivates these users to perform specific recoloring processes? Do they have precise expectations in mind when they begin recoloring a video? Can we measure their satisfaction with the final recolored results? What language do they use to communicate their motivations, expectations, and overall satisfaction?} Furthermore, contemporary academic research is often undertaken with a spin-out or start-up ethos in mind and particular focus is often placed the methods that are used to evaluate such outputs for commercialization in an HCI context; as they are often techniques that are shared with industry. This serves to outline the rationale for an academic spinout company, highlighting the strengths and weaknesses of the product and potential monetization of the intellectual property of the university.

To explore these questions concerning mobile video recoloring applications, we have developed a case-study based on a recent state-of-art recoloring technique~\cite{GROGAN2019} that explores the potential of such research in an emergent area of mobile software development. This study was undertaken to gain insight into how users with different levels of editing expertise use mobile recoloring applications and assess the measures outlined above to gain insight into the potential commercialization of one such product. In this paper, state-of-art image and video recoloring techniques are discussed (Sec.~\ref{sec::SOA}), the design of the recoloring application is described (Sec.~\ref{sec:colorTransferApp}), the methodology of the experiment is presented (Sec.~\ref{sec:userStudy}), the resultant data are gathered (Sec.~\ref{sec:results}), and finally, the conclusions that can be drawn are also discussed (Sec.~\ref{sec:discussion}).

\begin{figure}
    \centering
    \begin{minipage}[b]{0.48\columnwidth}
      \centering
      \centerline{\includegraphics[width=0.98\linewidth]{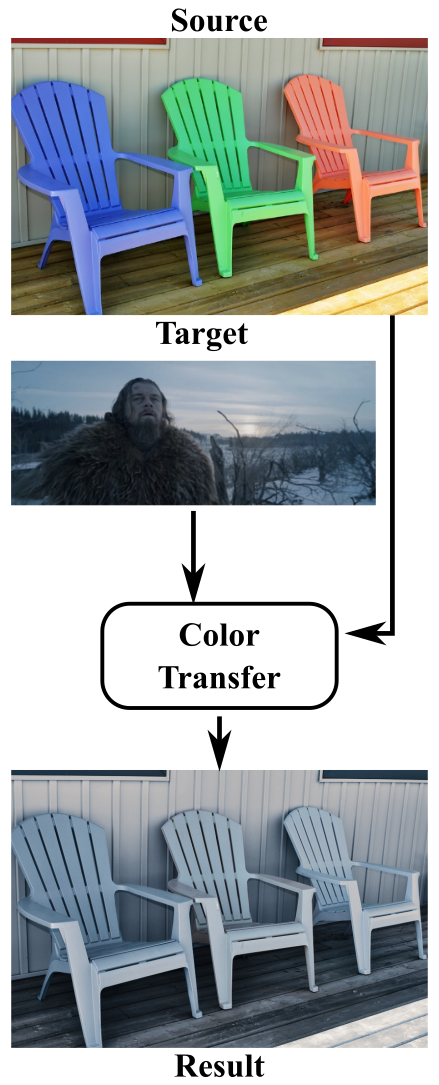}}
      \centerline{\small(a) Example-based transfer\normalsize}\medskip
    \end{minipage}
    \hfill
    \begin{minipage}[b]{.48\columnwidth}
      \centering
      \centerline{\includegraphics[width=.98\linewidth]{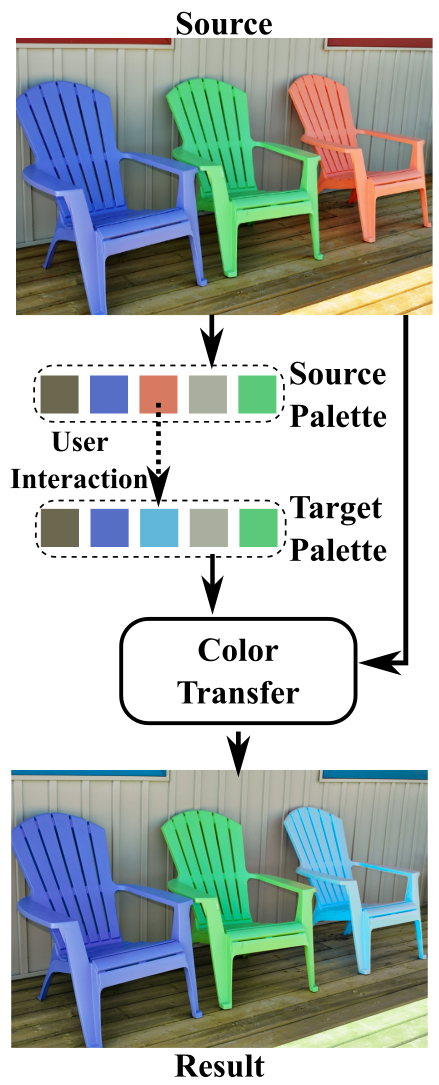}}
      \centerline{\small(b) Palette-based transfer\normalsize}\medskip
    \end{minipage}
    \vspace{-0.5cm}
    \caption{Examples of different color transfer techniques.}
    \label{fig:colorTransfer}
    \vspace{-0.4cm}
\end{figure}

\begin{figure*}[t]%[htb]
\begin{minipage}[b]{.19\linewidth}
  \centering
  \centerline{\includegraphics[width=.96\linewidth]{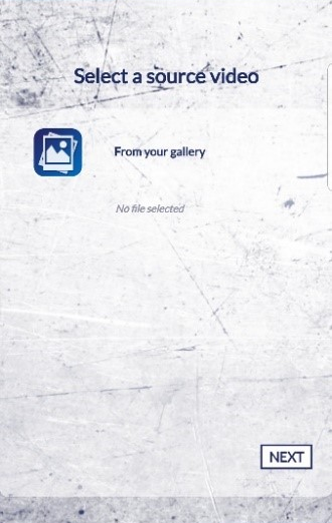}}
%  \vspace{1.5cm}
  \centerline{\small(a) Video selection.\normalsize}\medskip
\end{minipage}
\hfill
\begin{minipage}[b]{0.19\linewidth}
  \centering
  \centerline{\includegraphics[width=0.96\linewidth]{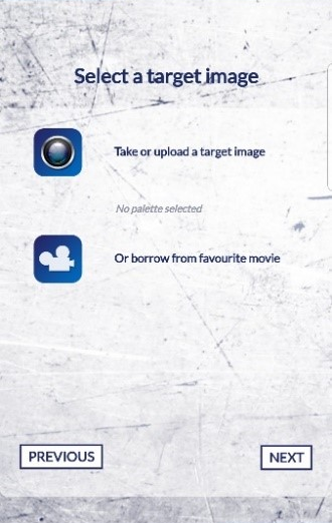}}
%  \vspace{1.5cm}
  \centerline{\small(b) Target image selection.\normalsize}\medskip
\end{minipage}
\hfill
\begin{minipage}[b]{0.19\linewidth}
  \centering
  \centerline{\includegraphics[width=0.96\linewidth]{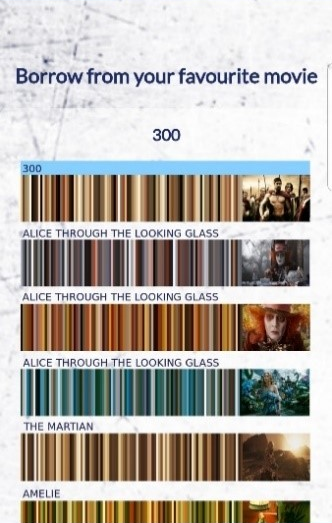}}
%  \vspace{1.5cm}
  \centerline{\small(c) Target images.\normalsize}\medskip
\end{minipage}
\hfill
\begin{minipage}[b]{.19\linewidth}
  \centering
  \centerline{\includegraphics[width=.96\linewidth]{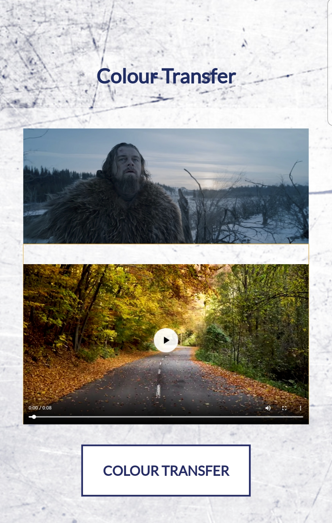}}
%  \vspace{1.5cm}
  \centerline{\small(d) Preview screen.\normalsize}\medskip
\end{minipage}
\hfill
\begin{minipage}[b]{0.19\linewidth}
  \centering
  \centerline{\includegraphics[width=0.96\linewidth]{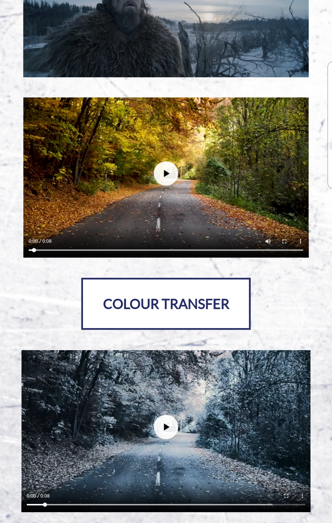}}
%  \vspace{1.5cm}
  \centerline{\small(e) Result screen.\normalsize}\medskip
\end{minipage}
\hfill
\vspace{-0.5cm}
\caption{The user interface of the mobile color transfer app.}
\label{fig:appUI}
\vspace{-0.4cm}
\end{figure*}

% ================================
%         State of the Art
% ================================
\section{State of the Art}
\label{sec::SOA}
Many different approaches to image and video recoloring have been proposed in the literature over the years. One popular approach is example-based image recoloring~\cite{Reinhard2001}, see \figref{fig:colorTransfer} (a), where an example image provides the color palette used to recolor the input image. Several example-based methods have since been proposed for both image~\cite{Pitie2005, Ferradans2013, GROGAN2019, Arbelot2017} and video recoloring~\cite{Grogan15,Frigo14,Bonneel2013}. The recoloring quality of these approaches has been assessed and user studies implemented to determine which example-based methods create the most appealing results~\cite{Hristova2015,GROGAN2019}. 
Similarly, palette-based image recoloring techniques have also become popular, see \figref{fig:colorTransfer} (b). These methods allow users to edit a small palette of colors to indicate what color changes they would like to see in their recolored images. Since these applications allow for more interactive experiences during the recoloring process, user studies have been undertaken to assess usability and quantify the overall user experience. Regarding novice users specifically, task-based analysis of recoloring an input image to match a target image have been completed~\cite{Chang2015}. Similar user studies have been presented to compare palette-based recoloring approaches~\cite{Zhang17} to the “replace color” functionality in Photoshop. Although these palette-based recoloring applications do not have the vast set of tools available in professional image and video editing software suites, for certain recoloring tasks they have been shown to provide adequate functionalities. 

In terms of video recoloring, while some methods estimate transfer functions per frame, with additional steps to account for temporal consistency~\cite{Bonneel2013}, many others simply extend image recoloring techniques to video data by applying the color transfer function estimated for a single frame to all frames of the video~\cite{Grogan15, Frigo14}. In this paper, the latter strategy is analyzed as it is faster to compute and does not generate any temporal inconsistencies in the resulting video. Again, while the quality of image recoloring techniques have been assessed through subjective experiments, their extensions to video data often have not \cite{GROGAN2019}. 

What is apparent from previous research is that there is still some research required to explore the motivations of users when choosing specific recoloring options~\cite{ghani1994task}, measure user satisfaction with outputs communicated through example-based and palette-based methods~\cite{chin1988development}, and discover whether this type of approach to video editing creates their desired recoloring results. Thus, we propose a case-study to assess user interactions with a mobile video recoloring application to determine user motivations, expectations, and satisfaction in terms of both the recoloring interface and the video results.

% ================================
%       Color Transfer App
% ================================
\section{Mobile Application for Video Color Transfer}
\label{sec:colorTransferApp}
In this case-study, a web-based Android application was developed as a prototype using a state-of-the-art video recoloring technology, see \figref{fig:appUI}. This example-based color transfer method was first proposed by Grogan, et al. \cite{Grogan151, GROGAN2019}. The underlying process allows the user to choose a predefined, example-based color distribution and transfer this distribution to an input image. An L2 based cost function was used to estimate a thin plate spline transfer function $\phi$ which transformed the color distribution of the input image to match that of the target %palette 
image, as in \figref{fig:colorTransfer} (a). This image recoloring method has been shown to extend to video by estimating a transfer function between a single frame of the input video and the input target %palette 
image, and applying the same transfer function to each frame of the video~\cite{Grogan15}. For this case-study, a random frame from the target video was selected from which to estimate the color transfer function $\phi$, and since the same function was applied to each frame, no temporal inconsistencies were created in the recolored video.

The application provided a front-end for this color transfer function, where users could select the video and target image %palette image 
combinations that they wanted to process. The target images are shown with a special color palette representation (see \figref{fig:appUI} (c)), which we refer to as ``palette'' or ``palette representation'' in this paper. Video recoloring was achieved via the following process. Upon opening the application, users are presented with an option to select a video for recoloring from the phone gallery (Fig~\ref{fig:appUI} (a)). Once an input video is selected, the user is moved to the second screen (Fig~\ref{fig:appUI} (b)) allowing them to select a target palette from the gallery or select a predefined target palette from within the application (Fig~\ref{fig:appUI} (c)). The predefined target palettes (12 in total\footnotemark) are frames from famous movies that have a strong color feel, for example, The Revenant, 300, Alice Through the Looking Glass, etc. Once a target palette has been selected and the user is happy with their input selections, they click ‘next’ and are then brought to the final option screen which displays the selected video and target palette together (Fig~\ref{fig:appUI} (d)). The user then clicks on the ‘color transfer’ button which begins the color transfer process and displays the final recolored video (Fig~\ref{fig:appUI} (e)). The source video and target palette are uploaded to a server in approximately 3.8 seconds; on average for a video with $960 \times 540$ resolution and a reliable (e.g., either WiFi or 4G without disruptions) internet connection. At this stage, the recolored video can also be downloaded by the user if required. The main aim of the application is to allow users to recolor their own videos so that they have similar look and feel as Hollywood films, or allow them to select their target image from their device.

% ================================
%           User Study
% ================================
\section{User Study}
\label{sec:userStudy}
To collect data concerning user motivations, expectations, and satisfaction, a controlled experiment was conducted. In this section, we discuss the methodologies applied.

\subsection{Apparatus \& Setup}
\label{subsec:testSetup}
The color transfer application was developed for the Android operating system, therefore, a Huawei P9 Plus smartphone (with $1080 x 1920$ display resolution) was supplied with the application pre-installed. A total of 16 videos\footnote[1]{More information regarding the selection procedure of the pre-selected source videos and target images is made available on the project website: \url{https://v-sense.scss.tcd.ie/research/vfx-animation/mobile-video-color-transfer-project/}. Interested readers are also referred to our recent work on exploring user-type opinions~\cite{grogan2020pilot}}. were placed on the device for the participants to select that encompassed a diverse set of contexts. All source video durations were 10 seconds. To speed up the recoloring process, the videos were resized to a resolution of $320 \times 180$ server-side before processing. This also sped up the time taken to download the processed video to the device. These videos ranged from drone footage to handheld cameras, all resembling personal videos taken for touristic purposes. Furthermore, users were asked to use the target palettes that were provided within the application, taken from famous movies, rather than using the option to upload target images from the phone's image gallery, see accompanying support materials for more details. This device set up ensured that we could identify what video and palette images were selected by each user for the reproducibility of results.

\subsection{Experiment Methodology}
\label{subsec:methodology}
Participants were recruited from within the Republic of Ireland. A call for participants was advertised online in December 2019, reaching out across the university, department, and research project network (i.e. mailing lists, social media accounts, etc.). The target participant pool was to be comprised of staff and students from the departments of engineering and computer science, as well as members of the university from outside of these schools. The experiment was performed over two weeks in January 2020. The study was conducted in a controlled laboratory environment to manage environmental continuities that may have affected the experiment, removing bias that may have been caused by differences in device, location, lighting, and internet connection. The experiment room was a well-lit, quiet space without windows. Upon arrival at the laboratory, participants were assigned a four-digit alphanumeric code to provide anonymity. Participants were then asked to report their age, gender, and occupation. At this stage, a user-cube approach was applied to identify user-types within the cohort~\cite{nielsen1993usability}. Accordingly, the participants were asked to self-identify their ability to use new technology, their familiarity with the video and image editing domain, and their expertise in the application and use of video and image editing applications on a fully labeled, 5-point Likert scale.

The main experiment comprised of three stages. The first stage of the experiment was to deliver a scripted demonstration on how to use the mobile phone and color transfer application (ensuring that all participants received the same instruction). During this time the participants observed how the color transfer application worked, they could watch the resulting recolored video results and could ask questions about completing the task for added clarity. 

\begin{table}[t]
\centering
    \caption{Questions in the post-task questionnaire. Open-ended questions were marked with a dot ($\cdot$).}
    \label{tab:questionnaire}
    \vspace{0.2cm}
    \footnotesize
    %\scriptsize
    \begin{tabular}{p{8.1cm}}  \hline \hline \noalign{\smalllskip}
    Question     \\ \noalign{\smalllskip} \hline \noalign{\smalllskip}
    1a) Why did you choose this palette? (choose one) \\ 
    ------ I like the colors \\
    ------ I want to change the mood of the source video \\
    ------ It complemented the source video \\
    ------ I want the video to look like the palette \\
    ------ It was different from the source video \\
    ------ I'm curious about how it will look \\
    ------ Random selection \\
    $\cdot$ Please tell us why you think this (optional) \\ \noalign{\smalllskip} \hline \noalign{\smalllskip}
    1b)	What do you expect the final video will look like? (choose one) \\
    ------ I have not thought about what the recolored video will look like \\
    ------ Any color mapping that matches the palette image well is in line with my expectations \\
    ------ I have a very specific idea of what I would like to see in the recolored video \\
    $\cdot$ Please tell us why you think this (optional) \\ \noalign{\smalllskip} \hline \noalign{\smalllskip}
    2) Did the results match your expectations? (choose one) \\
    ------ 5 point Likert scale - expectation\\
    $\cdot$ Please tell us why you think this (optional) \\ \noalign{\smalllskip} \hline \noalign{\smalllskip}
    3) How satisfied were you with the recolored video? (choose one) \\
    ------ 5 point Likert scale - satisfaction\\
    $\cdot$ Please tell us why you think this (optional) \\ \noalign{\smalllskip} \hline \noalign{\smalllskip}
    4) Overall, how appealing or unappealing was the recoloured video? (choose one) \\
    ------ 5 point Likert scale - appeal\\
    $\cdot$ a) Please tell us what you thought was appealing about the recolored video \\
    $\cdot$ b) Please tell us what you thought was unappealing about the recolored video \\ \noalign{\smalllskip} \hline \noalign{\smalllskip}
    5) Overall, how difficult or easy did you find this task? (choose one) \\
    ------ 7 point Likert scale - Single Ease Question \\
    \noalign{\smalllskip} \hline \hline 
    \end{tabular}
    \vspace{-0.4cm}
\end{table}

In the second stage, the participants were left alone and asked to use the application to complete video color transfer tasks. Participants were to remain seated and hold the smartphone at a comfortable distance from themselves without any specific limitations imposed. To control for novelty, the task was to be repeated a minimum of twice. All users used the same phone (Huawei P9 Plus) and the task was completed under the same network conditions throughout. Therefore, device and environmental conditions were maintained for all participants and there were no significant network-related delays reported. Once selected, both the video and target palette were uploaded to a remote server of the university where the color transfer function ($\phi$) was estimated and used to recolor the input video. Once recolored, the resulting video was sent back to the device and the recolored video displayed to the user. The total time from pressing ``color transfer'' to the download of the resulting video took approximately 8.6 seconds. During this time and before viewing the resultant video, the participants were asked to fill in a short questionnaire which included fully labeled, 5-point Likert scales and open-ended questions relating to their subjective motivations and expectations of the recolored video. After the color-transfer task had been fully completed and the participants had watched the resulting video, they were asked to fill out a brief user satisfaction questionnaire and repeat the experiment with a different set of parameters. In the third and final stage of the experiment, data were collected concerning the overall usability of the application. For this, the Single Ease Question (SEQ)~\cite{sauro2009comparison}, a 7-point rating scale, was used to assess how difficult the users found the task, see accompanying supplementary materials for more details. Each session took no more than 30 minutes. The full list of questions are given in Table~\ref{tab:questionnaire}.

% ================================
%           Results
% ================================
\section{Results}
\label{sec:results}
\subsection{Participants}
\label{subsec:participants}
\begin{figure}[t]
    \centering
    \includegraphics[width=\columnwidth]{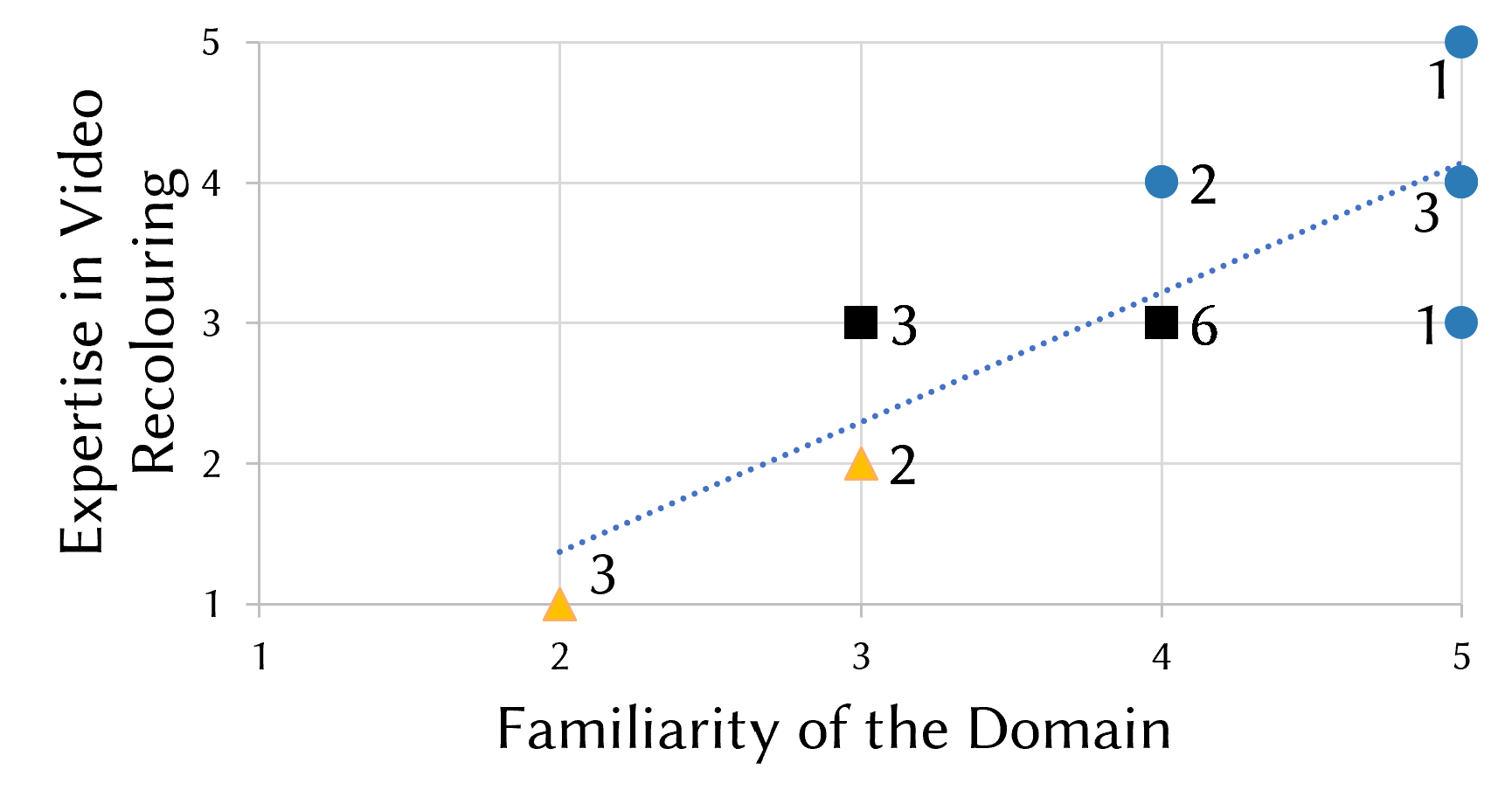}
    \caption{User-cube identifying user-groups: advanced (blue circles); end-users (black squares); novices (yellow triangles); numbers represent number of users.}
    \label{fig:expertise}
    \vspace{-0.3cm}
\end{figure}

A total of 21 participants volunteered to take part in the study (Male = 14; Female = 7), with an average age of 31.5 ($M = 31.50$; $SD = 5.50$). The employment sectors of the cohort consisted of students ($n = 10$), researchers ($n = 9$), and others ($n = 2$). Overall, the ability of the participant group to use new, digital technology was self-reported as being ``Good'' to ``Excellent'' ($M = 4.29$; $SD = 1.01$; $CI = 0.43$). The participants self-identified their familiarity with the video and image recoloring domain as being ``Moderately Familiar'' to ``Very Familiar'' ($M = 3.71$; $SD = 1.01$; $CI = 0.43$). When regarding their own expertise in the use of video and image recoloring applications, the participants rated themselves as being ``Average'' ($M = 2.95$; $SD = 1.07$; $CI = 0.46$). These three dimensions were used to identify individuals within the cohort as novices (users with some simple or mixed experiences with contemporary technology), primary and secondary end-users (users who are professionally motivated and routinely use this technology in their daily activities or who are motivated by occasional, contextual, or intrinsic interests), and advanced users (tech-savvy professionals with a broad knowledge of related activities and tools, a high degree of competency, and comprehensive expertise). All participants were sufficiently versed in the use of digital technology to contribute meaningful discourse on the topic and the participant group could be considered as containing broad perspectives from novices to advanced users who were discreetly familiar with the recoloring domain. From this data, user-groups were logically formed as novices ($n = 5$), end-users ($n = 9$); and advanced users ($n = 7$), see \figref{fig:expertise}.

\subsection{Motivations}
\label{subsec:motivations}
Participants were first asked to identify their motivations for choosing specific recoloring palette combinations. A selection of example responses was proposed as well as the opportunity to contribute personal opinions, see Table \ref{tab:ctq_1a}. Most users chose the option ``I’m curious about how it will look'' ($n = 14$), followed by ``I want to change the mood of the source video'' ($n = 10$). Users tended to have a specific motivation for picking their source video and target palette combination and did not use a random selection.

\begin{table}[t]
    \caption{Motivations for choosing a particular palette.}
    \label{tab:ctq_1a}
    \begin{tabular}{p{5.6cm} ccc}
    Answer                                                     & $T_0$ & $T_1$ & Total \\ \hline
    $\cdot$ ``I’m curious about how it will look''             & 7     & 7     & 14    \\
    $\cdot$  ``I want to change the mood of the source video'' & 6     & 4     & 10    \\
    $\cdot$  ``It complemented the source video''              & 3     & 4     & 7     \\
    $\cdot$  ``It was different from the source video''        & 4     & 2     & 6     \\
    $\cdot$  ``I like the colors''                             & 1     & 2     & 3     \\
    $\cdot$  ``I want the video to look like the palette''     & 0     & 2     & 2     \\
    $\cdot$  ``I chose at random''     & 0     & 0     & 0     \\
    \end{tabular}
\end{table}

User motivations relating to the chosen video and palette were then further contextualized via open-ended questioning. When asked to do so, participants reported on their intrinsic, subjective motivations. A content analysis (CA) was then conducted on this data to identify unique characteristics for choosing their recoloring arrangement via an inductive coding approach. A CA is a systematic research method used to determine the presence of themes or concepts~\cite{denzin2008collecting, adams2018qualitative}. The coding process was undertaken independently by two trained researchers. The intercoder agreement was measured as .80 or greater, which can be considered as acceptable for most scientific purposes. Assigned labels included: 36.67\% - changing the feel of the video to that of the palette (e.g. cold to warm, Autumn to Winter, make the video look older, make it look like a sunset, etc.), 23.33\% - the similarities or differences between the video and palette (e.g. make a scene look like it came from Game of Thrones), 13.33\% - personal preferences towards the colors of the palette (e.g. the complementary colors of the orange/teal in the palette), and other areas including 10\% - changing the quality of the video (i.e. brightness, saturation, darkness, etc.), and 10\% - creatively enhancing the video (making them more vibrant, colorful, and interesting, etc.) and 6.67\% - recreating a particular movie feel. When discussing their motivations to change the ``feel’’ of the original video, participants focused their comments on temperature, communicating this by using seasonal metaphors, for example:
\begin{quote}
    ``The video looks peak summer (lush green). The palette has yellow/dark wood shades. I want an autumn look.'' -- CT77 (advanced user).   
\end{quote}
These sentiments were also partially expressed by other participants, with users describing their desire to change the provided materials to be ``warmer'' or ``cooler'' concerning the color temperatures of the palettes displayed; using terms such as ``summer'' and ``winter'' as well as ``sunset'' and ``night'', to describe the differences between videos and palettes.
The participants reported that they were motivated to either complement their chosen video footage with a similar color palette or to juxtapose it with a visibly different palette, as one participant described:
\begin{quote}
    ``[I] chose the darkest palette to see what the snow would turn out like.'' -- YG89 (end-user).
\end{quote}
This included references to the color ``shades'', ``hue'', and ``light'' of both the source video and palette, for example:
\begin{quote}
    ``To see how the blue palette works on a pretty blue environment.'' -- SQ34 (end-user).  
\end{quote}
The cohort also made choices based upon their personal preferences towards color palettes that contained what was considered as being ``complementary colors'' and more simple partialities such as ``I like this palette very much''. Furthermore, personal motivations included opinions on the expected results of the recoloring process, for example:
\begin{quote}
    ``I believe both colors combined would make an interesting result.'' -- BN94 (end-user).   
\end{quote}

\subsection{Expectations}
\label{subsec:expectations}
\begin{figure}[t]
    \centering
    \includegraphics[width=\columnwidth]{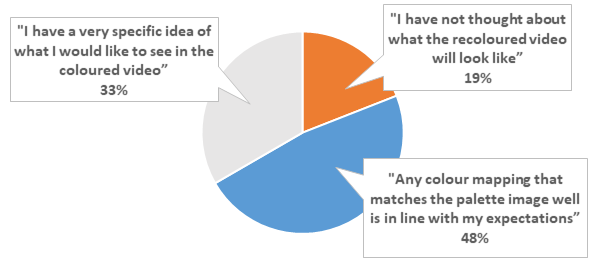}
    \caption{"What do you expect the final video will look like?"}
    \label{fig:pie}
    \vspace{-0.4cm}
\end{figure}

Participants were asked to describe and comment upon their expectations of the recoloring options they had selected. First, a selection of example responses was proposed, and the opportunity to expand answers further was then given, see \figref{fig:pie}. To begin, most users reported that ``Any color mapping that matches the palette image well is in line with my expectations'' ($n = 20$). Next was ``I have a specific idea of what I would like to see in the recolored video'' ($n = 14$). Finally, ``I have not thought about what the recolored video will look like'' ($n = 8$). The open-ended responses to this question were analyzed to further contextualize user expectations via a CA, as described previously. This analysis identified several areas of interest that linked user expectations to their palette choices, specifically, 40\% - to change the feel of the video and 30\% - to change the colors of the video. Lesser areas of interest were also identified, such as 10\% - recreating the feel of a particular movie, 5\% - highlighting the strong similarities/differences between palette and chosen video, and 15\% relating to other areas.

Participants expressed anticipation of changes to the feel of the original video in response to the chosen palette, specifically for their choice to create an effect seen in the palette visualization. This was verbalized via metaphors that related to seasonal changes and temperature, for example, ``I expected the natural scene to appear in colors resembling autumn'' and ``I think I will see enhanced blue shades which will give the video a cooler temperature''. This included expectations of ``natural'' color changes that resembled ``Summer'', ``Autumn'', and ``Winter'' scenes that enhanced the video with shades that altered color temperatures and the overall feel of the original video. More explicit color changes were expected that did not relate to the feel of the video but to specific color variations, for example, differences in ``brightness'', ``saturation'', and ``darkness''. These color-specific expectations had unambiguous outcomes and were communicated as ``I expect orange highlight and skin tone and teal-ish shadows'' and ``I expect the brighter colors on the balloon to be brighter''. Ultimately, the most common expectation in this category was to simply ``enhance'' the color of the original video.

\begin{figure}[t]
    \centering
    \includegraphics[width=\columnwidth]{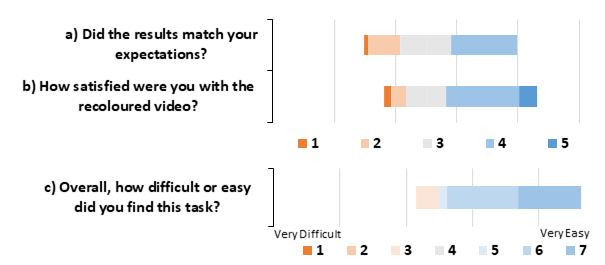}
    \caption{Results of the user's (a) expectations, (b) satisfaction, and (c) SEQ recolored video questionnaire.}
    \label{fig:graph}
    \vspace{-0.4cm}
\end{figure}

When video recoloring had completed and participants had viewed the new video, they were asked if the recoloring results matched with their expectations, see \figref{fig:graph} (a). The median scores for each user-type were measured as ``Met Expectations'' (Md = 3) over both trials. A Wilcoxon Signed Rank Test revealed no statistically significant differences in user expectations for $T_0$ and $T_1$ following the video recoloring process, $z = -0.468$, $p = 0.64$, with a small effect size ($r = 0.07$). Furthermore, at $T_0$ and $T_1$, a Kruskal-Wallis Test revealed that there was no statistically significant difference between user-types when matching user expectations, for $T_0$ $x^2 = (2, n = 21) = 0.434$, $p = 0.81$, and for $T_1$ $x^2 = (2, n = 21) = 3.02$, $p = 0.22$. Therefore, the user-type and number of trials did not affect overall user expectations. Further questioning expanded on why the participants chose their ratings. When exploring the qualitative user expectations data, the domain terminology used to describe the video recoloring results became much more focused on the resulting outputs. Users who gave a rating of ``Below Expectations'' commented that items such as shadows were too ``harsh'' compared to the original video, ``things got darker'', ``there were some artifacts, e.g. the yellow snow looked bad'', and ``[the] colors became saturated''. The users offered comments such as ``[the] red is a bit saturated'' and ``the video looked darker, more drained of color, not vibrant.''. These comments focused on the output achieved by the application, describing ``crushed blacks'', ``palettes'', and ``colors'', as well as ``saturation'', ``vibrancy'', and ``quality''. Users also continued to use temperature and seasonal metaphor to describe the outputted video, such as ``Autumn'', ``warm'', and ``natural''. Participants who rated their expectations higher supposed that the video looked ``more natural'' and ``[it] changed the mood of the video''. Other areas of interest extended to include references to moods, temporal coherence, and as one novice commented, ``The result looks much better (natural) than I expected''. Generally, users were happy that the recolored scene matched well, and that the mood of a video became more dramatic.

\subsection{Satisfaction}
\label{subsec:satisfaction}
Following questions on their expectations, participants were asked to rate their satisfaction with the recolored video. Participants reported that they were satisfied with the recolored videos with median scores for each user-type measured as being ``Very Satisfied'' (Md = 4) for all users over both trials, see \figref{fig:graph} (b). Exploring this data further, a Wilcoxon Signed Rank Test confirmed no statistically significant differences in user satisfaction for $T_0$ and $T_1$, $z = -0.291$, $p = 0.77$, with a small effect size ($r = 0.04$). Furthermore, a Kruskal-Wallis Test revealed that there was no statistically significant difference between user-type satisfaction, for $T_0$ $x^2 = (2, n = 21) = 1.12$, $p = 0.57$, and for $T_1$ $x^2 = (2, n = 21) = 4.86$, $p = 0.09$. When asked to explain their satisfaction further, participants reaffirmed that the recoloring was ``in line with what I imagined it would look” and ``it was like the idea I had in mind''. Reassuring comments cited that the ``[the] results were like what I had imagined'' and ``the video took on a different representation''. Other comments explored topics of ``color distribution and realism'', ``different representations of the story'', and as one participant expressed: % (see \figref{fig:samples}):
\begin{quote}
    ``If I were to do it manually, I would match the colors with the new palette exactly in the same way'' -- IS42 (novice).
\end{quote}
Positive measures of satisfaction focused on color changes that were ``expected'', ``[it] gave the scene a completely new look'', and that the ``overall result looks appealing''. Moreover, where satisfactory results were not observed, the participants stated that ``[I] would like to be able to tweak the results filter'', ``some parts of the video became too dark'', and ``It didn't match my expectations for the palette''. Satisfaction was not reached when the recolored video ``became too dark'', ``some details were lost'', and when the application ``negatively affected the quality of [the] video''. However, it was acknowledged by one end-user that ``The source video is difficult to recolor''.

\subsection{Task Difficulty}
\label{subsec:difficulty}
When assessing how difficult the recoloring task was, the SEQ revealed that all users considered the application to be easy to use ($M = 5.9$, $SD = 1.31$), see \figref{fig:graph} (c). The results of a Kruskal-Wallis Test indicated that there were no statistically significant differences in how difficult the users found the task, with $x^2 (2, n = 21) = 0.49$, $p = 0.78$. Therefore, all users found the task ``Easy'' to complete (Md = 6).

\section{Discussion}
\label{sec:discussion}
The recoloring mobile application task was appraised as being easy to complete by all users, supporting the existing technical abilities of advanced users, encouraging exploratory experiences for end-users, and using inclusive and clear communications for novices. We propose to explore this avenue of investigation in the future via in-depth usability driven evaluations

Participants were driven by specific motivations when recoloring the videos they were presented with. Changing the feel of a video was measured as being the most fundamental determinate for the cohort, communicated as temperatures and seasonal metaphors. This indicated that the prime motivations and inspirational triggers could be interconnected in an informal, creative language. Furthermore, the palette representations in the application were a source of inspiration that motivated the participants to explore the application in targeted, focused tasks that sought to recolor videos in the style of their favorite movies. Finally, personal preferences and internal motivations were also communicated as a driving force behind the cohort’s motivations; a factor that is difficult to account for without identifying user-types. However, changing the feel of a video and highlighting similarities or differences between the video and the palette are areas of the application that can be honed to suit user requirements.

The expectations of the users were observed as being closely related to their motivations, that is, in changing the feel of the final video. User expectations were also communicated as metaphors, meaning that the same temperature and seasonal metaphors could be used to describe the recoloring process. However, when it came to providing specific references to the changing of colors, more technical language was required. While user-type did not influence the satisfaction ratings of the cohort towards completed videos, the terminology used to describe satisfaction continued to be both metaphoric and technical. Therefore, when providing a clear, communicative language for describing recolored videos we recommend that developers first include metaphors as a motivational factor and secondly only provide technical terminology for tweaking settings in interactive applications that are aimed at a variety of user-types. Furthermore, the use of palette representation was confirmed as suitable for communicating the recoloring process to all user-types.

In terms of comments relating to the appeal of or user satisfaction with the result videos, we found that most negative comments related to unexpected changes in the brightness/darkness of the resulting video, or colors occasionally becoming oversaturated. When recreating several of the unsatisfactory result videos, it was found that these artifacts were due to the use of a single frame from the input video to estimate the transfer function $\phi$, which was then applied to all frames of the sequence. When palette images contained very bright, dark, or vibrant regions, with color values close to the boundary of the color space, colors that do not appear in the sample input frame but appear in other frames of the video were mapped outside of the color space, causing colors to appear crushed or overexposed. However, we found that this only happened when colors in the video changed dramatically across a video sequence. A regularization method, such as that proposed by Bonneel et al.~\cite{Bonneel2013}, could be implemented to overcome these issues, or a more sophisticated technique for selecting the source video frame, currently chosen at random, could be implemented. Negative comments towards satisfaction explored the loss of video quality after the recoloring process was performed. This was due to the video resizing applied before processing to speed up the video recoloring and download times. Additional steps to reduce computation times would remove the need for video resizing and allow the original video quality to be maintained. This would also allow for more instantaneous feedback of results, which was a feature that several users mentioned would be very beneficial during the recoloring. % process.

\section{Conclusion}
In this paper, we report on a case-study exploring aspects of an example-based color transfer application, with a focus on the users' experiences of motivation, expectation, and satisfaction of a potential spin-out application. The color transfer application enabled our users to use high-quality color graded content to shape their inputs. Overall, the color transfer application was found to be easy to use by all user types, regardless of their expertise and familiarity with the domain. We discovered that our users had intrinsic motivations in the form of basic curiosity or creative demands, which also engendered user expectations in most cases. Although our users' expectations might have varied depending on the color matching performance, most were satisfied with the results. Edge cases were largely due to mismatch between the input video and the selected color palette. We believe that these results can help develop better applications for content creation or improve satisfaction and user interfaces in similar recoloring applications. Furthermore, we believe that the findings can be used to report on the potential commercialization of such software as well as intellectual property creation and academic entrepreneurship.

\bibliographystyle{IEEEbib}
\bibliography{refs}

\end{document}